\documentclass[12pt,a4paper]{article}
\usepackage{amsmath}
\usepackage{amsthm}
\usepackage{amsfonts}
\usepackage{amssymb}
\usepackage{makeidx}
\usepackage[all]{xy}
\usepackage{verbatim}
\usepackage{xcolor}
\usepackage{enumerate}
\usepackage{textcomp}
\usepackage{braket}
\usepackage{hyperref}
\usepackage{graphicx}

\usepackage{mathtools}

\usepackage[sort&compress,numbers]{natbib}

\begin{document}
\sloppy

\title{\textbf{Thermal resonance in cancer}}
	
\author{Umberto Lucia $^{1,a}$ \& Giulia Grisolia $^{1,b}$\\ $^1$ Dipartimento Energia ``Galileo Ferraris'', Politecnico di Torino \\ Corso Duca degli Abruzzi 24, 10129 Torino, Italy\\  $^{a}$ umberto.lucia@polito.it \\  $^{b}$  giulia.grisolia@polito.it}

\date{\today}
\maketitle

\begin{abstract}
In the end of the second decade of 20th century, Warburg showed how cancer cells present a fermentative respiration process, related to a metabolic injury. Here, we develop an analysis of the cell process based on its heat outflow, in order to control cancer progression. Engineering thermodynamics represent a powerful approach to develop this analysis and we introduce its methods to biosystems, in relation to heat outflow for its control. Cells regulate their metabolisms by energy and ion flows, and the heat flux is controlled by the convective interaction with their environment. We introduce the characteristic frequency of a biosystem, its biothermodynamic characteristic frequency, which results to be evaluated by a classical heat transfer approach. Resonance forces natural behaviours of systems, and, here, we introduce it in order to control the fluxes through the cancer membrane, and to control of the cellular metabolic processes, and, consequently, the energy available to cancer, for its growth. The result obtained in some experiments is that the cancer growth rate can be reduced.


\end{abstract}
\section{Introduction}
Complex systems result as non-linear dynamical systems, composed by interacting subsystems, able to adapt to external perturbations of their environment \cite{ladyman}. Physics and chemistry of complex systems arose as an evolving interdisciplinary science from the theory of dynamical systems, in the 1960s \cite{Utham2020}. Physics and chemistry of complex systems allow us to modelling many phenomena such as self-replicating structures, non-equilibrium pattern formation, fluid dynamics, but also cancer growth \cite{wolfram,jiao}.

In biological and medical sciences, evolution is treated as a strategy of life at the level of the organism \cite{pienta1}, based on an interplay of genetic variation and phenotypic selection \cite{radman1}, because genes, and their variants, are selected in relation to their ability of encoding functions, useful to organism survival \cite{greaves1}. This last consideration is particularly true for cancer; indeed, cancer has been modelled as an adaptive system, based on natural selection, in order to allow any single cancer cell to become independent of its neighbours \cite{hanahan}.

The recent improvement of the complex physics approach to the analysis of cancer has pointed out that cancer is a complex adaptive
system \cite{pienta2,Utham2020}, as a consequence of some properties of it as  heterogeneous clonal expansion, replicative immortality, patterns of longevity, rewired metabolic pathways, altered reactive oxygen species,  evasion of death signals, metastatic invasion, etc. are all indications
of cancer's complex adaptive nature \cite{hanahan}. Indeed, the fundamental properties of the complex systems, but also of cancer, are non-linearity, emergence, self-organization, internal interconnection, etc. \cite{wolfram,gros}. In particular \cite{pienta1}:
\begin{itemize}
	\item Cells behave as agents: they are a set of active elements which interact in a selective way;
	\item Cancer cells activate some genes, turned off in
	normal tissue, in order to improve the characteristics useful to
	survive: this mechanism generates rules;
	\item Only the cells with similar adaptive mutations can survive: the components of the system gather together in relation to their similar abilities;
	\item Cancer behaviour is non-linear;
	\item Genetic instability allows cancer to fit easily and to expand.	
\end{itemize}

Consequently, a new viewpoint emerges in order to model organisms as highly regulated, complex, dynamic systems with meta-stability state around homeostatic levels \cite{bellavite1}. The meta-stability state is the result of fluctuations, amplifications and feedback cycles \cite{bellavite1} due to continuous oscillations of living systems between order and chaos, promoting survival. This meta-stability is the net result of continual
oscillations, rhythms, networks, amplifications and feedback cycles \cite{bellavite2,cramer}. In relation to oscillations, the phenomenon of resonance is well known in physics. Indeed, any system presents a proper oscillation frequency, and it can be forced to enter into vibration, if exited by a wave (mechanical or electromagnetic) at the frequencies close to its resonant one \cite{feynman1}.

From a thermodynamic viewpoint, a cell is an open system, able to convert its metabolic energy into mechanical and chemical works. The metabolic energy can be modelled as the heat inflow of a thermodynamic system. Consequently, cells can be modelled as a thermodynamic engine, which convert part of the inflow heat into work \cite{katchalsky}. In this context, normal and cancer cells presents two different cellular metabolism \cite{biochemistry}:
\begin{itemize}
	\item The Krebs cycle: a series of chemical reactions used by all aerobic organisms to release stored energy through the oxidation of acetyl-CoA derived from carbohydrates, fats, and proteins;
	\item The Warburg cycle:  a form of modified cellular metabolism found in cancer cells, which tend to favour a specialised fermentation over the aerobic respiration pathway that most other cells of the body prefer.
\end{itemize}
Indeed, in 1931, the Nobel laureate Otto Warburg showed that cancer cells, if compared with the normal ones, follow a different respiration pathway, which is characterized by a glucose fermentation, even when there is no lack of oxygen, highlighting how the variation on their metabolism was caused by a metabolic injury \cite{warburg}. Furthermore, the cytoplasmatic cells pH, and the extracellular environment, are directly linked to the cells membrane potential \cite{bingelli}. Comparing the polarization of quiescent cells with that of the differentiated ones, the latter result hyperpolarized \cite{becchetti}.

Any cell, as a thermodynamic engine, must outflow heat towards its environment \cite{katchalsky,whatslife,callen}, so, we expect that the two different cycles present two different heat outflow through the cell membrane. In order to model this process, we can consider an electric circuit equivalent for the cell membrane. But, in electric circuit, both transient and resonant phenomena can occur. So, we consider the possible equivalent behaviour in the heat transfer from the cell to its environment.

In this paper, we analyse the resonant heat transfer through the cancer cell membrane, and show how low frequencies electromagnetic waves can influence it, with a consequent decrease in the cancer growth.

\section{Materials and Methods}
Warburg showed the metabolic injury in cancer, pointing out the important role played by the energy conversion in biosystems \cite{warburg}. Cellular biochemical reactions convert external metabolites, considered as inflow of energy (inflowing heat for a direct thermodynamic cycle), into work (cell replication, protein synthesis, DNA and RNA transcription and translation, etc.), and wasted heat outflow towards cell environment \cite{katchalsky,whatslife}. Cells exchange energy and matter through their membrane \cite{caplan}, driven by the endogenous electric fields \cite{bustamante}.

Living cell membrane is a double lipid layer that separates the cytoplasm from the external environment. In membranes, some proteins perform a function of channels, across which the inflows and outflows of mass and ions can occur. It is usual to model the cell membrane as an electric RC circuit (Figure 1) \cite{rccircuit}.

\begin{figure}[h!]
	\caption{Electric analogy of a cell membrane. The cell membrane can be considered as a parallel RC circuit \cite{rccircuit}.}
	\vspace{0.3 cm} \includegraphics[width=\linewidth]{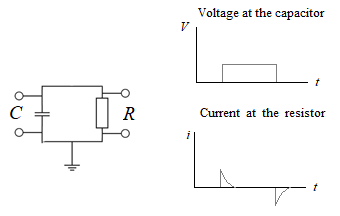}\label{RCcircuit}
\end{figure}

This kind of circuit presents both transient and resonant behaviour, in relation to the step or harmonic signal applied \cite{Horowitz,purcell}. If we consider the RC circuit, the transient behaviour of this circuit can be obtained in relation to the the current that flows across the resistor of resistance $R$ during the charge and the discharge of the capacitor \cite{Horowitz,purcell}:
\begin{equation}
i(t)=\frac{V_0}{R}\,e^{-t/\tau_{el}}
\end{equation}
where $i(t)$ is the current, $V_0$ is the value of electric potential applied to the capacitor, $R$ is the value of the electric resistance, and $\tau_{el} = RC$ is the characteristic time of the system. But, this characteristic time, related to the transient electric phenomenon, is also related to the resonant frequency; indeed, it results \cite{Horowitz,purcell}
\begin{equation}
\nu_{el} = \frac{1}{2\pi\,\tau_{el}}
\end{equation}

In relation to the heat transfer of the membrane, we can consider the thermo-kinetic lumped model. The cell exchanges heat power with its environment, remembering that the heat flux is related to its metabolism. This heat outflux occurs by convection with the fluids around any cell, and it results \cite{bejan-shape}:
\begin{equation}
\dot{Q}=\rho_{cell}Vc_{cell}\frac{dT_{cell}}{dt} = \alpha A (T_{cell}-T_{env})
\end{equation}
where $\dot{Q}$ is the heat power exchanged by convection, $\rho_{cell}$ is the cell mass density, $V$ is the volume of the cell, $c_{cell}$ is the specific heat of the cell, $T_{cell}$ is the cell temperature, $\alpha$ is the coefficient of convection, $A$  is the surface area of the cell, which varies during the phases of the development of the cell itself, and $T_{cell}-T_{env}$ is the temperature difference between the cell temperature and the environment temperature. As usually done in heat transfer, it is possible to obtain the characteristic time $\tau_{th}$ for the thermal transient \cite{bejan-heat}:
\begin{equation}
\tau_{th} = \frac{\rho_{cell}c_{cell}}{\alpha}\,\frac{V}{A}
\end{equation}
In analogy with the circuit model of the  cell membrane, we expect that there exists a resonant effect with a frequency  $\nu_{th} \approx 1/\tau_{th}$, with the hypothesis that $\nu_{el}=\nu_{th}$, because the electric circuit is only a theoretical model of the cell membrane.

So, if we irradiate the cancer by using an electromagnetic wave at this resonance frequency $\nu_{th}$, we expect to force the heat outflow from the cancer cell to the environment. Indeed, the heat power outflow of the equivalent electric circuit results:
\begin{equation}
\dot{Q}=RI_M^2\,\sin^2\big(2\pi\,\nu_{th}t\big)
\end{equation}
where $I_M$ is the maximum value of electric current in the equivalent circuit. But, at resonant state, the heat outflow is the maximum value of heat we can obtain. So, cancer decreases the energy available for some biochemical processes, such as differentiation, etc., with the consequence of decreasing its growth, because the increase of heat outflow makes the cancer cell less hyperpolarized, as it can be shown by considering the Nernst equation for the cell membrane \cite{katchalsky}:
\begin{equation}
\Delta \phi = \Delta G -2.3 \frac{R_uT_{env}}{F}\,\Delta \text{pH}=\Delta H -\dot{Q}\,\tau_{th} -2.3 \frac{R_uT_{env}}{F}\,\Delta \text{pH}
\end{equation}
where $\phi$ is the cell membrane electric potential, $H$ is the enthalpy, $R_u$ is the universal constant of gasses, $F$ is the Faraday constant, and $pH$ is the potential of hydrogen, and we have considered that the heat power outflow results $\dot{Q} = T_{env}\Sigma$, where $\Sigma$ is the entropy production rate in the environment and the heat results $Q=\dot{Q}\tau_{th}$.

In order to prove this result, we have developed some experiments, which confirm these results.

%
\section{Results}
Following the second law of thermodynamics, all the biochemical processes require energy, and any energy conversion process generates outflows of energy. Thus, it is possible to analyse the cells system behaviour, by following an engineering thermodynamic approach, considering the energy and mass balance. In cancer cells, an alteration on some processes related to energy and ion channelling have been shown, reducing their proliferation control. Heat transfer through cell membrane can be described by a thermo-kinetic lumped biophysical model. So, we can analyse the cell system as a black box, which is the usual approach used in engineering thermodynamics, considering all the internal biochemical reactions of the cell as the causes of the wasted heat outflow. The variation of the fundamental physical quantities, which control the biochemical reactions, can be controlled by controlling the heat transfer. Indeed, an electromagnetic wave at the thermal resonant frequency can force the heat transfer with a related change the membrane electric potential and the pH, conditioning the biochemical reaction and forcing them towards a normal behaviour.

The experimental proof of these theoretical results is shown in Table 1. It is possible to point out that:
\begin{itemize}
	\item The electromagnetic waves at thermal cell resonant frequency, reduce the growth rate of the cancer;
	\item The phenomenon is selective in relation to the frequencies used, as it must be for a resonant process.
\end{itemize}

\begin{table}[h!] \caption{Growth variation of some cancer cell lines after the exposure to the calculated resonant frequencies \cite{LUetaljtb2017,LUetalBBA2019}.}
	\centering
	\begin{tabular}{c c c }
		Cell line&Frequency&Growth variation\\
		&[Hz]&[\%]\\
		\hline
		A375P&$31$&$-15$\\
		HT-29&$24$&$-19$\\
		GTL16&$14$&$-24$\\
		MCF7&$5$&$-22$\\
		MDA-MB-231&$6$&$-18$\\
		SKBR3&$8$&$-18$\\
		\hline
	\end{tabular}
\end{table}

\section{Discussion and Conclusions}
The temperature difference between the inside and outside of any living cell is fundamental for the cell life, because this heat flow contributes to entropy variation with related reorganisation of the cell itself. The heat outflow, and the related entropy production, are caused by the biochemical and biophysical processes inside the cell. 

In this paper, we have developed the analysis of the thermal resonance of the cell membrane in relation to the heat exchanged by convection.

The results obtained highlight the role of the cell volume-area ratio, in relation to the heat fluxes control, with particular regards to the thermal resonant state of the living cell.

We have pointed out the existence of a proper time of response of any cell line to the heat exchange, as we expect in relation the the resonant phenomena. This time results related to the cells volume-area ratio, a geometrical parameter fundamental for the considerations on the fluxes and cells membrane electric potential variation.

Here, we have improved our previous results \cite{LUetalrsos2020,LUGGAAPP2020,LUGGEntropy2020} by focusing our analysis on the equivalent electric circuit model of membrane. This is a fundamental results, because it links our usual entropic analysis to the accepted model of membrane, in literature. In this way, we can explain the experimental results by linking together both the entropic analysis, developed in our previous papers, and the electric model of membrane, never considered before this paper. The results obtained by these different approaches converge to the same experimental results.

\bibliographystyle{report}
\bibliography{biblio}

\end{document}